\documentclass[aps,prb,preprint,groupedaddress]{revtex4}

 \usepackage{graphicx}
\newcommand{\be}{\begin{equation}}
\newcommand{\ee}{\end{equation}}

\begin{document}

\title{Three-body interactions involving clusters and films}

\author{Silvina M. Gatica$^1$, M. Mercedes Calbi$^1$, Milton W. Cole$^{1,3}$, and Darrell Velegol$^2$}

\affiliation{Departments  of
Physics$^1$ and Chemical Engineering$^2$, and Materials Research Institute$^3$, Pennsylvania State
University, University Park, PA 16802}

\date{\today}

\begin{abstract}

The three body (triple dipole) interaction of Axilrod, Teller and Muto (ATM) contributes 5 to 10 \% of the total energy of condensed phases of inert elements. It is shown in this paper for clusters and films that a much larger or smaller ATM contribution can arise for other geometries or other atomic species. The ratio $R$ of the three body interaction energy to the two body energy is evaluated for a wide variety of configurations. This ratio varies considerably with the geometry. For highly polarizable atoms in certain geometries, the magnitude of the three body energy is comparable to that of the two body energy and can be either attractive or repulsive. Systematic trends are established and explained.

\end{abstract}

\maketitle

\section{Introduction}

\begin{figure*}
\includegraphics[height=5in]{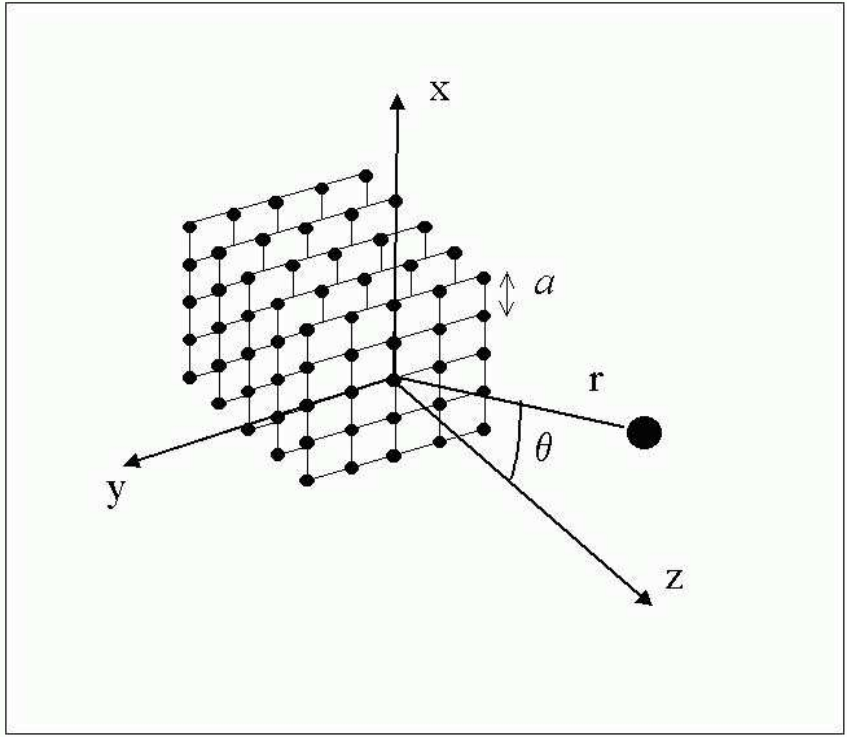}
\caption{Schematic picture of a cluster configuration of a five layer 
$s \times s$ square 
array, with $s = 5$ shown. The $z$ axis is perpendicular to the layers that interact with the external atom at {\bf r}.}
\label{confi}
\end{figure*}

The growing interest in problems involving interfaces has stimulated a
large  literature involving  computations  of the  properties of  such
systems \cite{book,davis}. Most of these studies have assumed that
the potential  energy of the system  is well approximated by  a sum of
two-body interactions  between pairs of the  constituent particles. In
some of  these instances, the pair  potentials are not  well known, so
the  implicit  or  explicit   neglect  of  many-body  interactions  is
appropriate, in view of the manifest quantitative limitations of these
particular calculations. In some other cases, however, the pair potentials
employed in the calculations have been constructed from empirical fits
to  properties  of  the  relevant  bulk  systems;  some  authors  have
suggested  that  such  potentials  inherently  include  higher  order,
many-body contributions \cite{milen,barker}. Implicit  in the latter point of  view is the
assumption that the many-body interactions in the given problem are of
a  similar  form and  magnitude  to those  in  the  bulk system.  That
assumption is evidently worth assessing in those cases where it can be
tested.
In this paper,  we explore the importance of  three-body van der Waals
(VDW) interactions
between a single atom
and
 clusters of  varying  sizes and shapes, from nearly spherical to flat.
In all cases, the atoms in the cluster are located at sites of a lattice
of lattice constant $a$, like that shown in Fig. \ref{confi}.
 Our study leads us to conclude
that  one should  {\it not} expect  three-body interactions  to  be properly
reckoned when an empirical pair potential is employed.

Our   analysis   of   the   three-body   energy  is   based   on   the
Axilrod-Teller-Muto (ATM) expression for that energy \cite{AT,M}. For 
specificity,
we  consider the case  when all  three atoms  are identical;  then the
interaction may be written:
\be
V_{ATM}({\bf r}_1, {\bf r}_2, {\bf r}_3)= \frac{9}{16}\frac{E_a \;
\alpha_0^3}{r_{12}^3 \; r_{23}^3 \; r_{13}^3} (1+3 \cos
\theta_1 \cos \theta_2 \cos \theta_3)
\ee
In this expression, the three angles represent the inner angles of the
triangle connecting the atoms (which  has sides of length $r_{12}$,
$r_{13}$ and $r_{23}$), the quantity $\alpha_0$ represents the
static polarizability of the atom and $E_a$  is an energy  (of order
the ionization  energy) characterizing
the assumed frequency dependence of the polarizability at imaginary frequency:
\be
\alpha(i\omega)=\frac{\alpha_0}{1+(\frac{\hbar \omega}{E_a})^2}
\ee
Representative  values of these parameters for  some atoms appear
in Table I.

The preceding ATM expression is a dispersion interaction, derived from third-order perturbation
theory, ignoring  exchange and  overlap of  the three  atoms' wave
functions; thus,  it is  known to be valid at large  separation, where
``large'' means relative to the atom's  hard-core interaction
diameter.  Empirical studies  have often  employed the  interaction at closer approach but there is some disagreement about the validity
of such applications \cite{say,lucas,bar1,bar2}; proposals have been presented that modify the behavior at small separation \cite{damp}. That important question is beyond the scope of the 
present work. The origin of the ATM interaction is not particularly intuitive. For that reason, among others, we discuss in Appendix A its origin in a particularly simple limit: the case when one atom (A) is far from the other two (B and C). In the limit when the characteristic frequency of A is much less than that of B and C, the three body (ATM) energy coincides with the mean electrostatic interaction between dipoles induced on B and C by the dipolar fluctuations of A.

The outline of this paper is the following. Section II presents a study
of the  VDW interactions  between an atom  and a monolayer  film, from
which we  make a straightforward extension  to include the  case of an
entire 3D crystal.  Section  III  describes  the  three-body  interactions
between an atom and a cluster of atoms. It will be shown in both of these
sections that the three-body energy $V_3$, relative to
the two-body  energy $V_2$,  is proportional to  the product
$n_s \alpha_0$  of the
number density of  the solid and the polarizability (Table 1). Because 
of this 
dependence, we focus our attention on a dimensionless ratio 

\be
R=\frac{V_3}{V_2} \; \frac{1}{n_s \alpha_0}
\ee

$R$ depends  very   much  on   the  specific
geometry and can be either positive or negative. In  particular, $R$ has  
a drastically different
dependence on separation in the two problems investigated here. In the
cluster case,  the ratio decreases  with distance (vanishing  at large
distance)  while in  the  atom-solid case,  the magnitude of $R$ increases  with
distance, asymptoting to a constant value that may be derived from the
continuum  theory of  VDW  interactions. For  reference,  we state  the
corresponding result in  the case of the cohesive  energy of inert gas
solids; the ratio of three-body to two-body energies then is typically
less than  10 per cent \cite{bar1,bar2}. This small value reflects in part a 
significant cancellation of ATM terms of opposite sign in a bulk material. The 
present study finds similar cancellation in some geometries, but not others. 
In the latter case, the three-body energy represents an important term that 
should be included in calculations of system properties. Section IV 
discusses the  implications of
this work.

\section{Three-body atom-monolayer and atom-solid interactions}

It was  shown by  Dzyaloshinskii, Lifshitz and  Pitaevski \cite{DLP} that the
long range  interaction between  a nonpolar atom \cite{dip} at  position {\bf r}=(0,0,z)
and a semi-infinite solid, that occupies the half-space $z<0$, assumes the form

\be
V({\bf r}) \sim -\frac{C_3}{z^3}
\ee

\be
C_3=\frac{\hbar}{4 \pi} \int \; d\omega \; G(\omega) \; \alpha(i\omega)
\ee

Here $G(\omega)$, the surface-response  function at imaginary  frequency, is
determined from the dielectric function $\varepsilon(i\omega)$

\be
G(\omega)=\frac{\varepsilon(i\omega)-1}{\varepsilon(i\omega)+1}
\ee

Among  the  questions  addressed  in  this  section  is  the  relative
importance  of   contributions  to  $C_3$  from   pair  interactions
and
three-body interactions; we call  these contributions $C_3^{(2)}$
and $C_3^{(3)}$, respectively. We  first  consider  the  problem  
of an  atom  situated  at  position
{\bf r}=(0,0,z)  interacting with  a  square  lattice ($s \times s$) of
atoms  that
comprise a  monolayer film, in the $z=0$ plane;  its crystalline
lattice directions are  oriented along the $x$ and  $y$ axes. 
In  this  case,  the  two-body  van  der  Waals
interaction may be written:

\be
V_2 = -\sum_i \frac{C_6}{r_i^6}
\ee

Here, the sum is  over all atoms ($i$) of the film,  each at distance
$r_i$ from the external atom, and the coefficient of the asymptotic VDW pair
interaction is

\be
C_6=\frac{3}{4}\,E_a\,\alpha_0^2
\ee

\begin{figure*}

\includegraphics[height=4in]{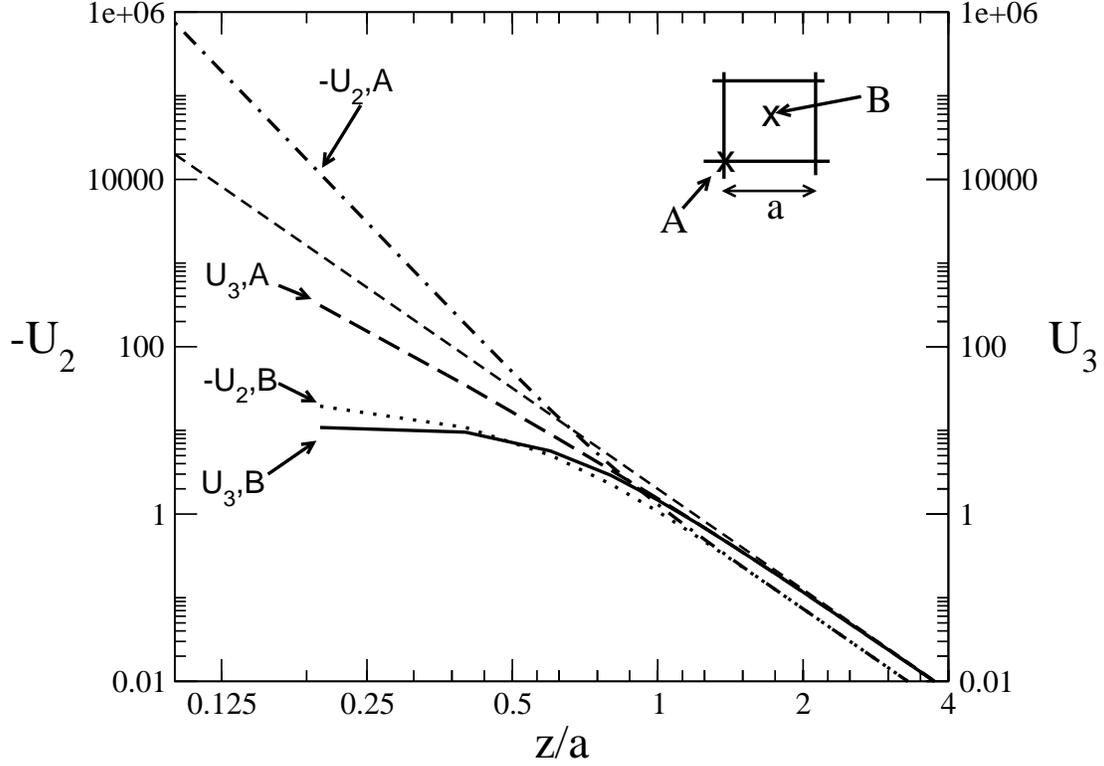}
\caption{Dimensionless two-body and three-body interactions 
$U_2$ and $U_3$ (defined in Eqs. (9) and (10)) for the case of an atom 
above a 60 $\times$ 60 monolayer film as  a function  of
perpendicular distance $z/a$, for two indicated values of the lateral 
position of the  atom: above  an  atom of  the  film, A-top 
and above  a midpoint  of  the square 
lattice  of  the  film, B. The short-dashed line corresponds to the 
asymptotic $z^{-4}$ prediction of Eq. (11).}
\label{v23}
\end{figure*}

This  expression  (first  obtained  by  London \cite{london}) is 
derived  from  the
frequency-dependent polarizability, Eq. (2). For $s=60$, Fig. \ref{v23} presents the dimensionless two-body potential

\be
U_2=\frac{V_2}{\epsilon_2} \;\;\; ; \;\;\; \epsilon_2=E_a \,(\alpha_0 n_s)^2
\ee

\noindent Similarly, we define 

\be
U_3=\frac{V_3}{\epsilon_3} \;\;\; ; \;\;\; \epsilon_3=\epsilon_2 \,\alpha_0 n_s
\ee

\noindent Values of $\epsilon_2$ and $\epsilon_3$ are presented in Table I. One  observes a significant dependence  of the
energy on lateral position only at small $z/a$, since at $z>a$ the
film looks  continuous.  At large distance,  one finds for $V_2$ a
result that  can be  derived from an  integration over  the monolayer,
which has a two-dimensional (2D) density $1/a^2$:

\be
V_2({\bf r}) \sim - \frac{C_{mono}^{(2)}}{z^4} \;\;\;\;\;\;\;;\;\;\; C_{mono}^{(2)}=
\frac{\pi}{2}\,\frac{C_6}{a^2}
\ee

The  superscript (2)  in the  coefficient indicates  that  this energy
originates  from the  pair interactions,  i.e.  Eq. (7).  Also shown  in
Fig. \ref{v23}  is the  total  three-body  energy,  obtained by  summing  ATM
interactions  between all  pairs of  atoms  in the  monolayer and  the
external atom (at {\bf r}):

\be
V_3({\bf r}) = \sum_{i<j}\; V_{ATM}({\bf r}_i,{\bf r}_j,{\bf r})
\ee

\begin{figure*}

\includegraphics[height=4in]{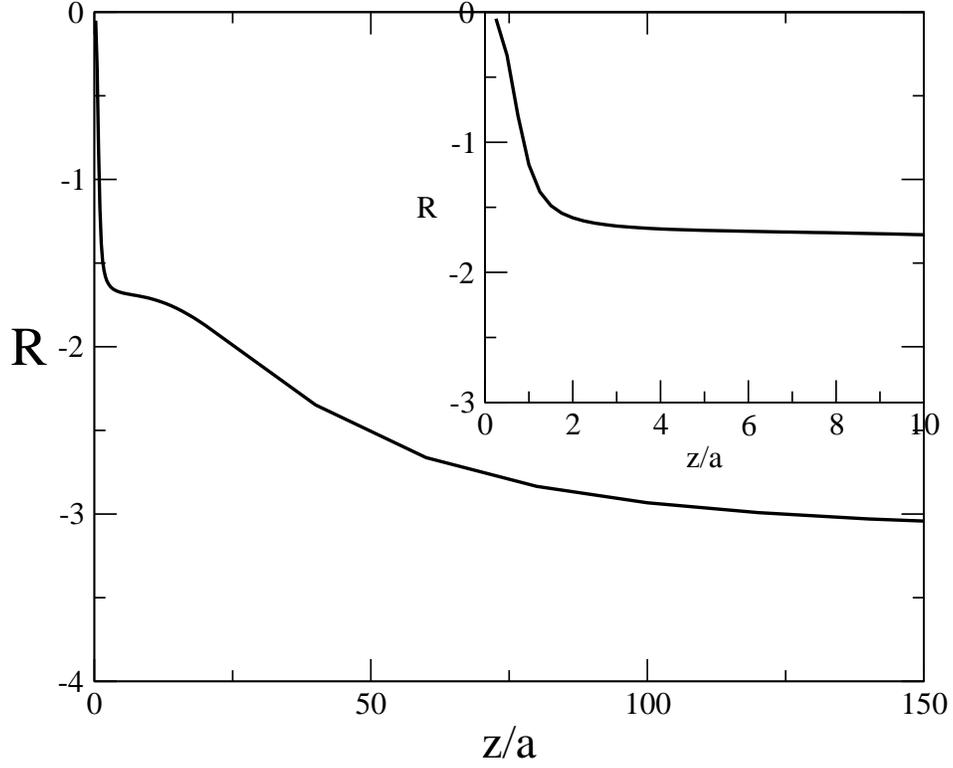}
\caption{Ratio $R$, defined in Eq. (3),  of the three-body and two-body interaction
between an atom at position (0,0,z) and a 60 $\times$ 60 monolayer film, as a function of the perpendicular distance $z$.
In the inset is shown an enlargement of the region of small $z < 10a$.}
\label{mono}
\end{figure*}

This energy  is positive  because the contribution of the repulsive  terms in the  sum exceed  those of attractive  terms.
In Fig. \ref{mono} we show the ratio $R$ (Eq. (3)) for this case. 
One observes  in the 
inset of this  figure  that at a distance $z$ 
much larger than
the  lattice constant $a$ (but smaller than the monolayer's lateral extent) 
the ratio of these
interactions is $R_{monolayer} \approx -1.6 $.
We will derive this result below.
For distances larger than the size of the monolayer (equal to $60 a$ in the case of the Fig. \ref{mono}), the value of $(V_3/V_2)$ approaches a finite value that depends on the
size of the monolayer. This value is discussed in the next section.

Now, let us  address the problem of the  three-body interaction in the
case of an atom above a half-space solid, consisting of layers located
at  $z=-na$,  where $n=0,1,2,...$  First,  we  evaluate the  two-body
contribution  by summing  the monolayer  energy over  all  layers.
  We note, from  the Euler-Maclaurin theorem
and Eq. (11), that the asymptotic dependence is 

\be
V_2({\bf r}) \sim - C_{mono}^{(2)} \sum_n \frac{1}{(z+na)^4} \sim
 -\frac{\pi}{6} n_s \frac{C_6}{(z-\frac{a}{2})^3}
\ee

This  leads  to the  result  that  the  two-body contribution  to  the
asymptotic potential of Eq. (4) involves a coefficient

\be
C_3^{(2)} = \frac{\pi}{6} n_s C_6
\ee

We  now  evaluate  the  analogous three-body  asymptotic  coefficient,
$C_3^{(3)}$. The expression is obtained from Eq. (12), with
the sum of  the  ATM  interactions taken  over pairs  of  atoms
throughout  the half-space $z \leq 0$. To determine this quantity requires
consideration of two-kinds of terms in the sum, leading to:

\be
V_3 = V_3^{inter} + V_3^{intra}
\ee

The interlayer term  $V_3^{inter}$ represents the three-body 
contribution arising  from pairs of solid atoms
coming from different
planes, while  the intralayer  term $V_3^{intra}$ involves
pairs  of atoms  from the
same plane. Fig. \ref{inter-intra} shows the resulting $R^{inter}$ and
$R^{intra}$ terms.

In Appendix B, we prove that the first term in  Eq. (15) is identically
zero at large  separation, leaving just the second  term. We note that
the  ratio of  this surviving  intralayer term  to the  analogous pair
interaction term was calculated above  for a single layer, as shown in
Fig. \ref{mono}; the ratio asymptotes to a constant value mentioned 
above. This
means that, in the semi-infinite solid  case, the same ratio applies 
to contributions
from all planes, leading to  exactly the same asymptotic ratio for the
solid and the monolayer. Thus, asymptotically,

\be
\left(\frac{V_3}{V_2}\right)_{solid} \sim \frac{C_3^{(3)}}{C_3^{(2)}}
\sim -1.6  n_s \alpha_0
\ee

We now explain why this
numerical result  arises (and could have been  anticipated without any
numerical  calculations). To  do so,  we employ  the Clausius-Mossotti
relation:

\be
\frac{\varepsilon(i\omega)-1}{\varepsilon(i\omega)+2}=\frac{4 \pi}{3}
n_s \alpha(i \omega)\equiv \xi(i \omega)
\ee

Assuming that the quantity $\xi$ is  small (the case of small $V_3/V_2$), 
we may expand the relation for
the asymptotic VDW coefficient, Eq. (5):

\be
C_3=\frac{\hbar}{4 \pi} \int \; d\omega \; \alpha(i\omega) \left[\frac{3}{2}
 \xi{(i\omega)}\left(1-\frac{\xi(i\omega)}{2}....\right)\right]
\ee

\be
C_3=C_3^{(2)}+C_3^{(3)}....
\ee

The  identification  of successive  terms  with  increasing number  of
interacting particles is physically plausible and is demonstrated here
for the first two terms. If one inserts into these equations the Drude
expression for  the polarizability, Eq. (2), assuming the  same species
for external atom and solid atoms, one obtains

\be
C_3^{(2)} = \frac{\hbar}{2} n_s \alpha_0^2 \int \, d\omega\,
\frac{1}{\left[1+(\frac{\hbar \omega}{E_a})^2\right]^2}
\ee

\be
C_3^{(3)} = -\frac{\hbar \pi}{3} n_s^2 \alpha_0^3 \int \, d\omega \,
\frac{1}{\left[1+(\frac{\hbar \omega}{E_a})^2\right]^3}
\ee

The integrals may be evaluated, leading to values of the two relevant
contributions to the dispersion coefficient:

\be
C_3^{(2)}= \frac{\pi}{8} n_s \alpha_0^2 E_a = \frac{\pi}{6} n_s C_6
\ee

\be
C_3^{(3)}= -\frac{\pi^2}{16} n_s^2 \alpha_0^3 E_a
\ee

\noindent The ratio of coefficients results:

\be
\frac{C_3^{(3)}}{C_3^{(2)}}=-\frac{\pi}{2} n_s \alpha_0
\ee

This  result is fully  consistent with  the behavior found numerically,
with coefficient $ R=-\pi/2 \approx$ -1.6, Eq. (16).  This confirms our attribution  of the physical  origin of the
first two terms in the expansion of $C_3$.

We also explore the dependence of $V_3$ on the orientation of the monolayer. This is shown in
Fig. \ref{mono-angles}. The result is that $V_3$ depends very strongly on the orientation, becoming
 attractive when the single atom and the monolayer are coplanar. The origin of this behavior will be explained in the next
  section.

\section{Three-body interaction between a single atom and a cluster}

Now, we address the problem of a single atom at a distance $z$ from a 
quasi-spherical
cluster. By ``quasi-spherical'' we mean a cluster of $N$ atoms constructed 
from the closest $N$ points to the center of a simple cubic lattice of sites. 
The resulting ratio $R$ in this configuration is shown in Fig. \ref{cl-at} 
for different sizes and orientations of the cluster.
There are at least three particularly interesting dependences seen in this figure. 
The first is that $R$ goes asymptotically to zero, in contrast to 
the nonzero constant value in the case of a monolayer at large 
distance. The second  is the presence of a maximum in the magnitude of $R$. A third is that $R$ depends 
strongly on the orientation of the cluster. For example, in the case 
of the  clusters with 7 and 33 atoms, the three-body interaction is 
attractive for $\theta=0$ but repulsive for large values of 
$\theta$. In the other two cases shown in this figure (clusters with 
19 and 93 atoms), the trend is reversed.

We now explain why $R$ vanishes at large $z$ for quasi-spherical 
clusters.  The ATM expression Eq. (1) at large $z$ is simplified because 
two of the three interatomic distances are essentially equal to $z$ 
and two of the interior angles of the three-atom triangle sum to 
$\pi$, while the third angle nearly vanishes. Then, the energy may be 
written

\be
V_{ATM}(z)=\frac{C}{z^6} \; \sum_{i<j} \; \frac{(1 - 3 \cos^2 \phi_{ij})}{r_{ij}^3}	\;\;\;\;\;  ; \;\;
C=\frac{9}{16} E_a \alpha_0^3
\ee

\noindent Here, the sum is over pairs of atoms in the cluster, separated by 
distance $r_{ij}$ and $\phi_{ij}$ is the angle between their separation vector 
${\bf r}_{ij}$ and the vector connecting the atoms to the external atom. Since the term in parentheses can vary between (-2) and 1, the sign of the resultant sum is not obvious. However, in a case  where the separation vectors are randomly oriented, the average value of cos$^2 \phi$  is 1/3. Similarly, for an {\it infinite} simple cubic lattice, the average is 1/3, as may easily be established (for any crystalline orientation). In either of these cases, therefore, the sum in Eq. (25) vanishes identically. In the case of a finite quasi-spherical cluster, the ATM sum is nonzero, due to residual surface contributions, but it is small compared to the two-body energy, which is proportional to the number of atoms in this large separation limit. Thus, it is expected that $R$ should be very small for symmetric clusters at large distances.

To understand the intriguing angular dependence in Fig. 6, we undertake additional calculations, as follows. We consider the long range interaction for two geometries shown in Fig. 7. In Fig. 7(a) appears the interaction for anisotropic clusters consisting of ($s$=1,2 or 3) rows of atoms, with $N$ atoms per row, oriented parallel to the $z$ axis, on which the atom is located. One observes for $s=1$ row that $R$ is positive ($V_3 <0$) for all values of $N$. This behavior can be understood readily from Eq. (25) because  $\phi_{ij} =0$ for all pairs within the cluster. Hence, every term is negative and the sum increases essentially linearly with the number of nearest neighbor pairs (and hence $N$) for $z/a >> N$. Cases involving $s > 1$ yield somewhat less attractive ATM sums, relative to two-body sums, as indicated in the $s=2$ and 3 curves in Fig. 7(a). To summarize, needle-like clusters, pointed toward the external atom, have negative ATM interactions, enhancing the two-body attraction.

What about more symmetric structures? In Fig. 7(b), we show results for systems consisting of atoms located at the sites of $N$ parallel squares of size $s$. The distance from the external atom to the first square is
    $z=10 s$, and $R$ is plotted as a function of $N/s$. We see that for $N/s<1$
    the 3-body contribution is repulsive, while for $N/s>1$ it is attractive.
    In the case $N/s=1$, that corresponds to a cubic configuration, the ratio $R$ is almost zero at
    this distance. Thus, the asymptotic behavior of the cube is analogous to that of the small
     clusters. We note from Fig. 7 that this vanishing is an ``accident'', in that nonzero values of opposite signs occur for $N > s$ or $N < s$.

Now suppose, instead, that our system  consists of atoms on sites of 
three parallel squares, as in Fig. \ref{trilayer}. As expected, if 
the side of the squares is exactly 3, $R$ goes to zero, while the 
asymptotic value of $R$ becomes finite (and negative) for $s>3$. Note that this trend is consistent with the asymptotic value $R \approx -3.1$ for one layer, with $s=60$, shown in Fig. 3, and can be understood by the following argument. At large $z$ ($\gg sa$), the film atoms are esentially equidistant form the external atom. Hence, the two-body energy is $V_2=-NC_6/z^6$. As for the ATM energy, in the case of normal incidence at large $z$, the term involving 3 cosines in Eq. (1) vanishes, leaving:

\be
V_{ATM}=\frac{9}{16} E_a \, \alpha_0^3 \, \Sigma
\ee

\noindent The sum $\Sigma$ is over all pairs of atoms; for a film of very large extent, this may be written

\be
\Sigma=\sum_{i<j} \frac{1}{r_{ij}^3}\equiv N_a \frac{k}{a^3} \;\;\; ; \;\;\; (N_a \gg 1)
\ee

\noindent Here, $N_a$ is the total number of atoms and $k=4.7$ is a numerical factor evaluated from the sum. Since $C_6=(3/4) E_a \alpha_0^2$, the ratio $R$ becomes

\be
R= -\frac{3}{4} k \approx - 3.5
\ee

The large $z$, large $s$, limit of Fig. 3 approaches this value. For large $z$, but $s$ not very large compared to 1, the contribution of sites on the boundary of the film causes $\Sigma$ to be less than $Nk/a^3$, so the ratio $R>-3.5$ (as seen in Fig. 8). This is a significant reduction in magnitud for finite $s$ because the fraction of sites on the border of an $s \times s$ square is $4 (s-1)/s^2$.
 
These results lead to a relatively simple and comprehensive picture 
of the role of three-body interactions. Lines of atoms pointing toward 
the external atom lead to attraction, while planes of atoms 
perpendicular to the cluster-atom separation vector lead to repulsive 
interactions overall (see Appendix A). Note, in particular, that the explanation 
clarifies the peculiar angle-dependence in Fig. 6. 
In the process of making the clusters nearly spherical,
those with $N=7$ and 33 end up with single atom prominences on their external 
"faces". The $N=7$ cluster consists of an atom at the origin and one along each of the +/- axes. The $N=33$ cluster is composed by a cube of side 3 plus an additional atom at the center of each face. In particular, when $\theta=0$ these configurations mimic the effect of a line of atoms pointing toward the external atom originating an attractive interaction as indicated by the dashed lines in the left panels of Fig. 6. Instead, the other two clusters with $N=19$ or 93 face the external atom with an almost complete plane of atoms perpendicular to the cluster-atom separation vector, leading to repulsive interactions at small angle $\theta$ (dashed lines in right panels of Fig. 6).

\section{Discussion and final remarks}

The calculations in this paper use the Axilrod-Teller-Muto 
three-body interaction to supplement the contribution from two-body 
calculations
of van der Waals dispersion forces. Specific calculations reveal that
material geometry - both the morphology and orientation of clusters - is
important, and that changing the orientation of a cluster can cause the
overall VDW potential to change from attractive to repulsive.  This is
especially true for clusters with high polarizability.

        Our calculations suggest three heuristics for predicting van der
Waals dispersion interactions between an atom A and a material B (e.g.,
surfaces, clusters, or atoms) interacting by a force along a unit 
vector $\hat{e}$:  1) If B has a large dimension perpendicular to 
$\hat{e}$, this
tends to produce a repulsive 3-body force with A, thus reducing
the total dispersion attractions; 2) if B has a large
dimension parallel to $\hat{e}$ (i.e., depth), this tends to produce attractive
3-body forces with A, thus increasing the total dispersion
attractions; 3) if B is quasi-spherical, this tends to decrease the magnitude of 3-body forces with A, meaning that the pairwise interaction sum
gives a good approximation to the total interaction.

        This work has produced other important conclusions as well.  For
example, in assessing 3-body interactions between an atom and a
semi-infinite solid, only the intra-plane contributions to the 3-body
effect are important; that is, distinct layers of the solid do not
interact significantly with the external atom to produce 3-body ATM effects.  In addition,
analytical calculations show that the 3-body effect asymptotes to zero
for quasi spherical clusters, as opposed to semi-infinite materials.  

        At least two interesting questions arise from this work.  First, since we
did not assess 4-body interactions, it remains to be seen whether 4-body
interactions are of the same order of magnitude as 3-body interactions.
Presumably such interactions scale as one higher power of the expansion parameter $n_s \alpha_0$. Second, the effect of coating layers on particles can be examined by a
similar methodology. In 2-body effects only, the coating layer might
not alter the VDW forces significantly; however, based on the present
work, the 3-body forces could produce even repulsive net interactions.

\begin{acknowledgments}
We are grateful to L.W. Bruch for stimulating and helpful discussions.
This research has been supported by the National Science Foundation, the Ben Franklin Technology Center and the Environmental Protection Agency. 
\end{acknowledgments}

\appendix
\section{Nature of the ATM interaction at large distance}

In this section we explore the microscopic nature of the ATM interaction
 by deriving its form (in a particular case) using a distinct 
{\it alternative} to the conventional approach to deriving the 
three-body interaction. The method described here provides a simple 
interpretation to this energy that is consistent with the general 
notion that such dispersion interactions arise from interactions 
between fluctuating dipole moments. 

To be specific, we consider the situation when two atoms (B and C) are 
close together while the third atom (A) is located at large distance 
from the pair. Furthermore, for convenience, we make a specific 
assumption about the values of their characteristic energies that is 
different from that used to derive Eq. (1). The latter assumes that 
all three energies are the same; here, instead, we assume that $E_A$, 
the 
characteristic energy of atom A, is small compared to the energies 
(both $E'$) of B and C, meaning that A has a slowly fluctuating 
dipole moment. In this limit, we evaluate the triple dipole 
interaction by setting it equal to the electrostatic interaction 
between the two instantaneous dipoles (on B and C) that are induced 
by dipolar fluctuations of the third atom, A.

Assume that the external atom (A) is located at the origin and the 
B, C pair of 
atoms is located at distance $z$ on the z axis. A dipole moment fluctuation ${\bf p}$ on A will produce an electric field 
${\bf E}=[-{\bf p}+3p_z{\hat{\bf z}}]/z^3$ at the neighboring positions 
of the B, C pair, where $\hat{\bf z}$ is the unit vector in the z direction . 
Atoms B and C polarize instantaneously (in comparison with A), 
so that they develop an identical induced dipole moment 
${\bf p}_{induced}= \alpha_0 {\bf E}$. The electrostatic interaction 
energy between two such dipoles is then obtained in the usual way, 
resulting in:

\be
\langle E_{BC} \rangle = \frac{\alpha_0^2}{z^6}\;\langle{\bf p}^2\rangle _A \frac{(1 - 3 \cos^2 \phi_{BC})}{r_{BC}^3}
\ee

Here, we have have  averaged over the time-dependent fluctuations 
of ${\bf p}$ on atom A, with the average denoted by brackets. The 
angle $\phi_{BC}$ is that between the BC separation vector and the AB separation 
vector. Note that this energy is usually nonzero, varies as the inverse 
cube of the B-C separation and the inverse sixth power of the distance 
to A and has an interesting angle dependence. To explore the 
comparison with Eq. (1), we evaluate $\langle {\bf p}^2\rangle _A$ within the Drude 
approximation \cite{fano}, Eq. (2). The result is $\langle {\bf p}^2 \rangle _A = (3/2) E_A \alpha_0$. 
Hence, we arrive at the expression for the three-body interaction 
in this approximation:

\be
V(z) = \frac{3}{2}\; E_A \;\frac{\alpha_0^3}{z^6} \;\frac{(1 - 3 \cos^2 \phi_{BC})}{r_{BC}^3}
\ee

With one exception, this equation coincides with the result obtained 
from Eq. (1)  in the present geometry  (two angles summing to $\pi$ and 
the third vanishing). The exception is the value of the numerical coefficient, which is here (3/2) instead of (9/16) in Eq. (1). This difference 
arises because Eq. (1) is obtained from the ATM model for a case when 
the characteristic energies coincide, while Eq. (A1) assumes that the 
energies are very different, $E_A \ll E'$. If this latter set of energies were 
inserted in the {\it general} ATM expression, it would yield the 
same (3/2) coefficient found here. Thus, we have identified a plausible 
and intuitive explanation (induced dipole interactions) of the origin 
of this ATM interaction and shown that the energy computed in the 
present limit agrees with ATM formula in the case of the same 
interacting atoms.

\section{Interlayer three-body interaction}

The total interlayer ATM interaction for an atom above a half-space 
crystal may be written as a  sum of terms in
which the external atom ($0$) is fixed in position and one sums over all
pairs  of  internal  atoms.  We  evaluate the  sum  by  considering  a
particular individual atom ($i$) within the solid and summing over atoms
in all layers  below it, which we represent  as a continuum. McLachlan
evaluated the  many-body interaction for precisely this  geometry \cite{mc1,mc2},
depicted in Fig. 9. Atom $i$ lies at distance $b$ (of order a lattice
constant) above the continuum while the external atom lies at
distance $h=z + na + b$ above this continuum; here $n$ is the layer number
for atom $i$. The atoms labelled $0$ and $i$ are separated  laterally by a
distance $R$. McLachlan's resulting expression for this many-body 
interaction energy involves two coefficients:

\be
C_{s1}=\frac{3 \hbar}{\pi} \int \; \xi(i\omega) \,\alpha^2(i\omega)\, d\omega
\ee
\be
C_{s2}=\frac{3 \hbar}{\pi} \int \; \xi^2(i\omega) \,\alpha^2(i\omega)\, d\omega
\ee
Since  the first  of these  coefficients is  proportional to  the third
power  of the  polarizability,  it represents  a  contribution to  the
three-body energy  explored in this  paper. The $C_{s2}$ coefficient  is of
fourth order in the polarizability, as discussed  explicitly below. Corresponding  to these
coefficients, the  expression of McLachlan yields  a limiting behavior,
for $h>>b$, as follows:
\be
V_M= C_{s1} U_1(R,z) + C_{s2} \;\frac{1}{(R^2+h^2)^3}
\ee
\be
U_1(R,z)= \frac{1}{(R^2+h^2)^3}\left[\frac{4}{3}- \frac{2 h^2}
{R^2+h^2}\right]
\ee

Note  that the  $U_1$ term  is negative  for $R/h < 2^{-1/2}$
and  positive for
larger  $R/h$,  with  a  maximum  at  $R=h$. Hence, there  tends  to  be
cancellation between these regimes of distance. To determine the total
contribution from  all of the  atoms in layer  $n$, we need to sum
this expression for $V_M$ over a monolayer  of atoms; for $h>>b$, we
can achieve this by integrating this term over all positions $\vec R$ in the plane of atom $i$. Note that
\be
\int \; U_1(R,z) \, d^2\vec R = 0
\ee
We conclude  that  the   three-body  contribution   from  interlayer
interactions  vanishes  at  large  distance. The  remaining  interlayer
contribution is fourth order, which is not considered in this paper. For 
completeness, however, we evaluate this contribution from layer
$n$, of two-dimensional density $1/a^2$:
\be
V_{4n}= \frac{1}{a^2} \int \; \frac{C_{s2}}{(R^2+h^2)^3} \; d^2 \vec R =
-\frac{\pi}{2}\frac{C_{s2}}{a^3}\frac{1}{h^4}
\ee
We may  then integrate this contribution  over all of  the layers ($n$),
yielding
\be
V_{4inter} = - \frac{\pi}{6}\frac{C_{s2}}{a^2}\frac{1}{z^3}
\ee
The ratio of this term to the two-body term is found to be
\be
\frac{V_{4inter}}{V_2}=\frac{C_{s2}}{C_6} = \frac{5}{8} \left(
\frac{2 \pi \alpha_0}{a^3}  \right)^2
\ee

This is smaller  by a factor $\sim \alpha_0/a^3$ than  the
corresponding ratio in Eq. (14), for the three-body  intralayer
term. We note, however, that we
have  not considered  fourth order  intralayer terms,  
which  might be larger and/or of opposite sign.

\begin{table}[tbh]
\begin{tabular}{|c|c|c|c|c|c|} \hline
$\;\;$Atom$\;\;$ & $\;\;\alpha_0$ (\AA$^3$)$\;\;$ &$\;\;\;\;\alpha_0
 n_s\;\;\;\;$ & $\;\; E_a$ (eV)$\;\;$ &$\;\;\;\epsilon_2$ (eV)$\;\;\;$& $\;\;\;\epsilon_3$ (eV)$\;\;\;$ \\ \hline \hline
$^4$He&0.205 & 0.0045  &  24.59 & 5$\times 10^{-4}$ & 2$\times 10^{-6}$  \\ \hline
Ar   &1.64   & 0.041   &  15.76 & 0.026           & 0.001              \\ \hline
C    &1.76   & 0.176   &  11.26 & 0.35            & 0.06     	 \\       \hline
Xe   &4.04   & 0.067   &  12.13 & 0.05            & 0.004	\\	\hline
Si   &5.38   & 0.269   &  8.15	& 0.59            & 0.16         \\	\hline
Au   &5.8    & 0.342   &  9.23	& 1.1             & 0.37       \\	\hline
Pt   &6.5    & 0.179   &  8.96	& 0.29            & 0.05      \\	\hline
Pb   &6.8    & 0.225   &  7.42	& 0.38            & 0.08      \\	\hline
Ag   &7.2    & 0.422   &  7.58	& 1.4             & 0.57      \\	\hline
Cu   &7.31   & 0.617   &  7.73	& 2.9             & 1.8       \\	\hline
Al   &	8.34 & 0.502   &  5.99	& 1.5             & 0.76      \\	\hline
Fe   &8.4    & 0.706   &  7.87	& 3.9             & 2.8       \\	\hline
Mg   & 10.6  & 0.457   &  7.65	& 1.6             & 0.73      \\	\hline
Na   &  23.6 & 0.600   &  5.14	& 1.9             & 1.1       \\	\hline
Ca   &25     & 0.582   &  6.11	& 2.1             & 1.2       \\	\hline
K    &43.4   & 0.576   &  4.34	& 1.4             & 0.83      \\	\hline

\end{tabular}
\caption{Static polarizability ($\alpha_0$), polarizability times number density ($\alpha_0 n_s$) and ionization potential ($E_a$) of various atoms and corresponding solids (values taken from Ref. 7). The energies $\epsilon_2=E_a (\alpha_0 n_s)^2$ and $\epsilon_3=\epsilon_2 (\alpha_0 n_s)$ are the scale units for $V_2$ and $V_3$ used in Fig. 2.} 
\end{table}


\newpage

\begin{figure*}
\includegraphics[height=4in]{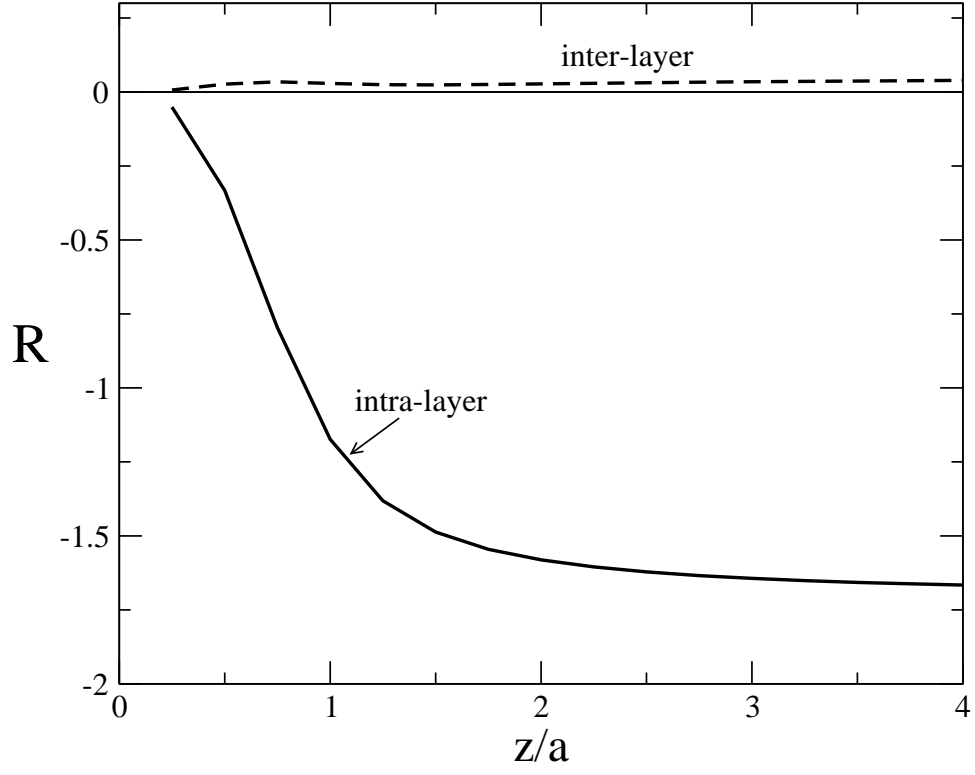}
\caption{Inter-layer and intra-layer contributions to the ratio $R$ for an atom above a bilayer film.}
\label{inter-intra}
\end{figure*}

\begin{figure*}
\includegraphics[height=3in]{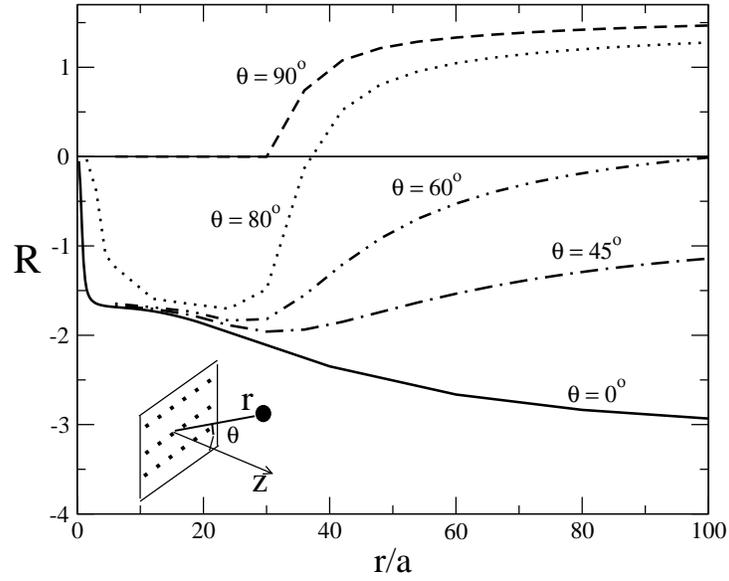}
\caption{Ratio $R$ as a function of the distance $r$ to the center of the 60 $\times$ 60 monolayer for different orientations $\theta$.}
\label{mono-angles}
\label{mono-angles}
\end{figure*}

\begin{figure*}
\includegraphics[height=4in]{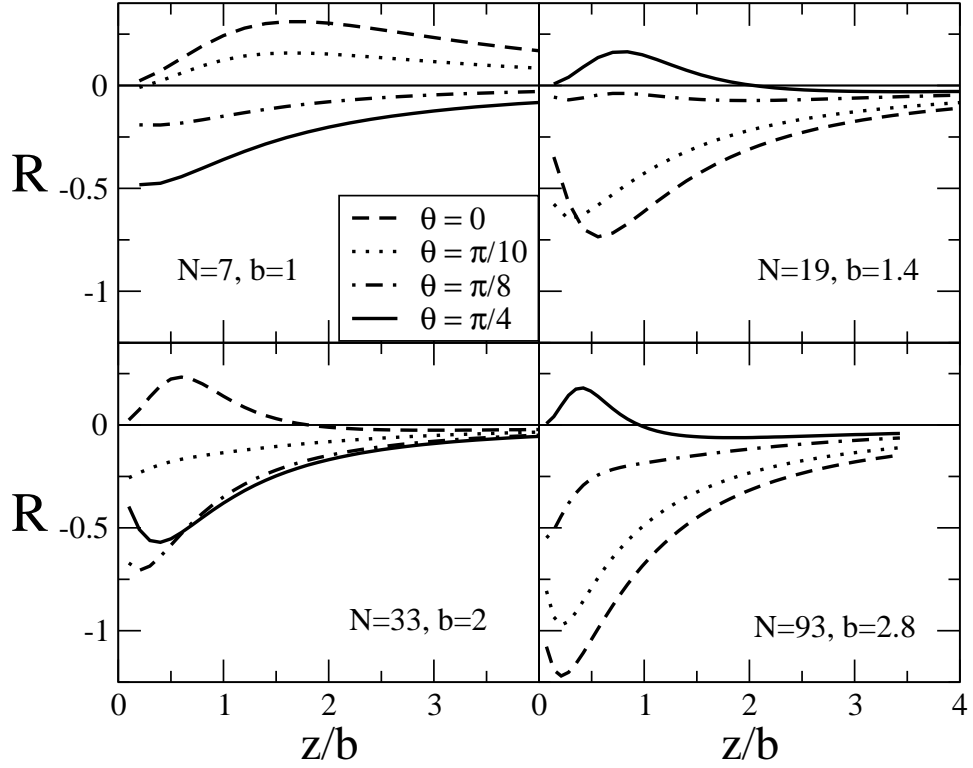}
\caption{Ratio $R$ of the three-body to two-body interactions between an atom and a quasi-spherical cluster of radius $b a$ as a function of the distance to the surface of the cluster,  for different orientations $\theta$. $N$ is the number of atoms in the cluster.}
\label{cl-at}
\end{figure*}

\begin{figure*}
\includegraphics[height=3in]{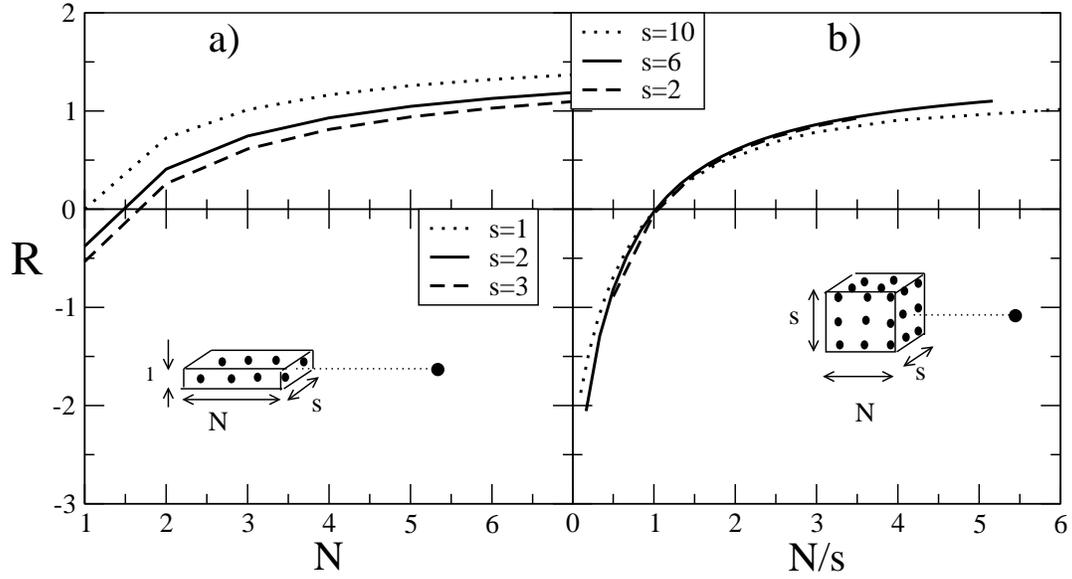}
\caption{(a) Ratio $R$ evaluated at a distance $z=10a$, for the case of an atom and a  cluster of atoms in one line, two lines and three lines,
as a function of the number of atoms in the line, N. (b) Ratio $R$, evaluated at a distance $z=10s$, for the case of an atom and a  cluster of atoms of dimension $s \times s \times N$.}
\label{lines}
\end{figure*}
\newpage
\begin{figure*}
\includegraphics[height=4in]{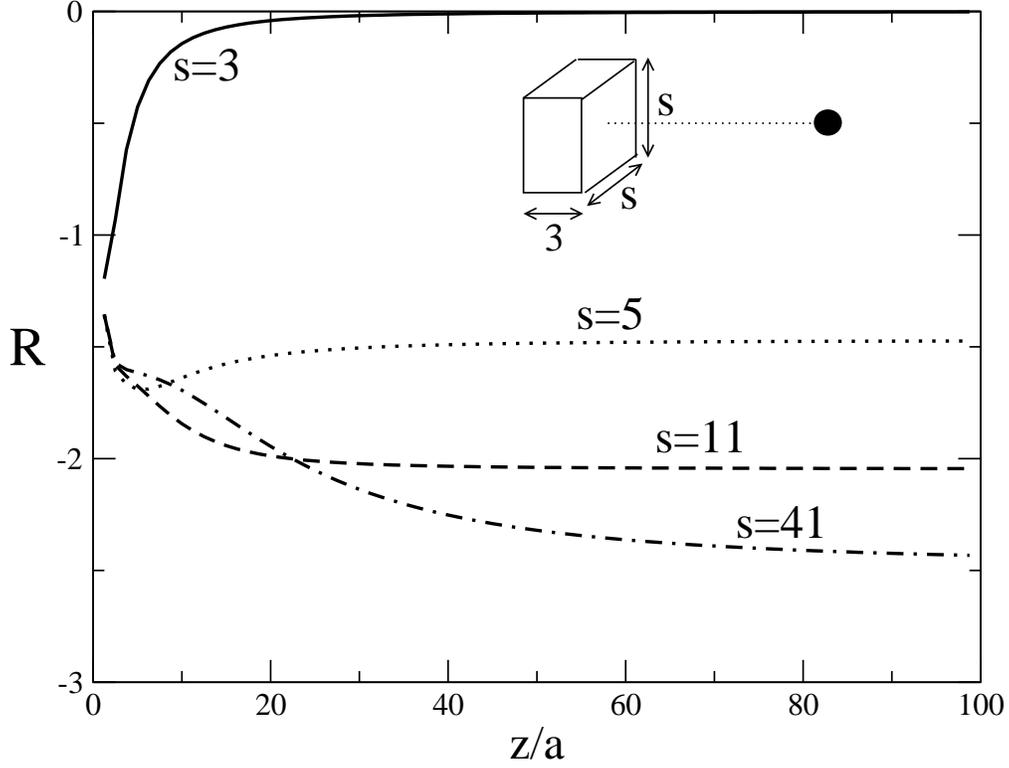}
\caption{Ratio $R$  for the case of an atom and an $s \times s \times 3$ 
cluster of atoms, as a function of the perpendicular distance $z/a$.}
\label{trilayer}
\end{figure*}

\begin{figure*}
\includegraphics[height=3in]{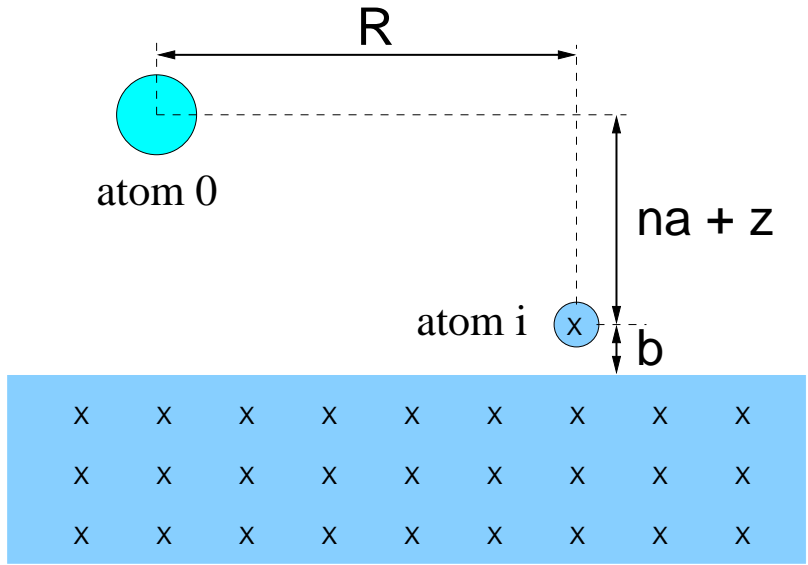}
\caption{The geometry corresponding  to atoms 0 and i above a half-space, whose
interaction is mediated by the solid below. The crosses represent the lattice structure of the solid.}
\end{figure*}

\end{document}